\documentclass[12pt]{iopart}
\usepackage{epsf}

\begin{document}
\title[A permanent magnetic film atom chip for Bose-Einstein
condensation]{A permanent magnetic film atom chip for Bose-Einstein
condensation}

\author{B.V.~Hall, S.~Whitlock, F.~Scharnberg\footnote[1]{Present address: \textit{Institut f\"{u}r Quantenoptik, Universit\"{a}t Hannover, 30167 Hannover, Germany}}, P.~Hannaford and
A.~Sidorov}

\address{ARC Centre of Excellence for Quantum-Atom Optics and
Centre for Atom Optics and Ultrafast Spectroscopy,
Swinburne University of Technology, Hawthorn, Victoria 3122,
Australia}

\ead{brhall@groupwise.swin.edu.au}

\begin{abstract}

We present a hybrid atom chip which combines a permanent magnetic
film with a micromachined current-carrying structure used to realize
a Bose-Einstein condensate. A novel TbGdFeCo material with large
perpendicular magnetization has been tailored to allow small scale,
stable magnetic potentials for ultracold atoms. We are able to
produce $^{87}$Rb Bose-Einstein condensates in a magnetic trap based
on either the permanent magnetic film or the current-carrying
structure. Using the condensate as a magnetic field probe we perform
cold atom magnetometry to profile both the field magnitude and
gradient as a function of distance from the magnetic film surface.
Finally we discuss future directions for our permanent magnetic film
atom chip.
\end{abstract}

\pacno{39.90.+d,03.75.Be,39.25.+k,07.55.-w}

\section{Introduction}

A recent technological advance in the area of quantum degenerate
gases has been the development of the {`}atom chip{'}.  These
devices exploit tightly confining, magnetic potentials, created by
low power current-carrying wires to simplify the production of
Bose-Einstein condensates \cite{Han01,For01}.  In addition, they
provide the freedom to realize intricate magnetic potentials with
features of size comparable to the atomic de Broglie wavelength.
Atom chips have been used to realize atomic waveguides and transport
devices for Bose-Einstein condensates \cite{Lea02,Mul99}.   These
tools allow controllable manipulation of ultracold neutral atoms,
with potential applications in quantum information processing
\cite{Cal99,Wang05} and atom interferometry \cite{Shi05,Sch05a}.

For current-carrying wire-based atom chips, technical limitations
are imposed by current noise and spatial fluctuations in the current
density leading to increased heating rates and fragmentation of cold
clouds \cite{For02,Est04}.  In addition, near-field thermal noise in
conductors is responsible for a fundamental atom loss mechanism
\cite{Jon03}.  Atom chips incorporating permanent magnetic materials
are expected to overcome many of these difficulties.  These
materials offer the possibility of ultra-stable magnetic potentials
due to their intrinsically low magnetic field noise. Moreover
permanent magnetic films are thin and relatively high in resistance
compared to current-carrying wires, properties which strongly
suppress thermal magnetic field noise \cite{Sch05b}.  Permanent
magnetic materials with in-plane magnetization have recently been
used to demonstrate trapping of cold atoms \cite{Bar05} and to
produce BEC on a magnetic videotape \cite{Sin05a,Sin05b}. Here we
employ a novel magnetic material with perpendicular anisotropy,
developed specifically for applications with ultracold atoms.
Perpendicularly magnetized materials allow arbitrary 2D patterns to
be written in the plane of the film and provide magnetic field
configurations analogous to those produced by planar microfabricated
wires \cite{Eri04,Jaa05}.

 In this paper we report the realization of a
permanent magnetic film/machined conductor atom chip which has been
used to produce a $^{87}$Rb Bose-Einstein condensate (BEC).  In
section 2 we present a simple model for a thin film of
perpendicularly magnetized material which results in straightforward
equations for the magnetic field near the edge of the film. We then
describe the principle of trapping ultracold atoms in the potential
formed by the permanent magnetic film (the film trap).  TbGdFeCo
materials are then introduced in section 3 with a description of the
deposition process. We measured the bulk properties of the magnetic
film using both a superconducting quantum interference device
(SQUID) and a magnetic force microscope (MFM). Section 4 describes
the construction of the ultra high vacuum (UHV) compatible atom
chip. This includes a current-carrying structure to provide time
dependent control of surface based potentials and is used to form a
conductor-based magnetic trap (the wire trap).

The apparatus and experimental procedures used for making a BEC
independently with the film trap or the wire trap are described in
section 5. In section 6 we apply the BEC as a novel ultracold atom
magnetometer by measuring the spatial decay of the magnetic field
from the film. High precision trap frequency measurements in
conjunction with radio frequency output coupling also allow the
direct determination of the associated magnetic field gradient. In
conclusion we speculate on future directions for our permanent
magnetic film atom chip.

\section{Simple model of a permanent magnetic film}

\begin{center}
\begin{figure}
\begin{center}
\epsfbox{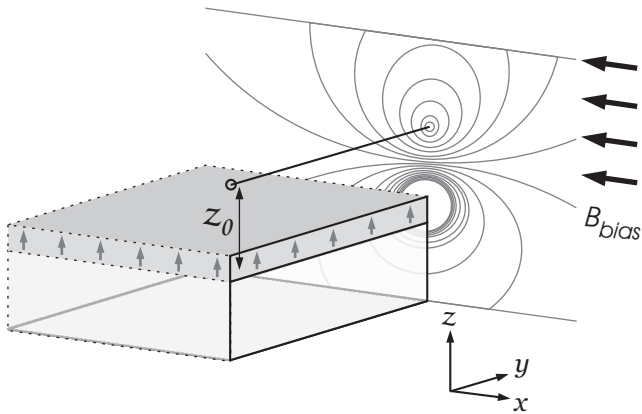}
\end{center}
\caption{\label{film}A simple model describes the magnetic field of
a semi-infinite, perpendicularly magnetized thin film in combination
with a uniform bias magnetic field.}
\end{figure}
\end{center}

Consider a semi-infinite rectangular magnet with magnetization $M$
and thickness $h$ (Figure~\ref{film}). The magnet lies in the $xy$
plane with one edge aligned along the $y$ axis. The magnet is
uniformly magnetized in the $+z$ direction and the distance from the
film is large with respect to the film thickness ($z~\gg~h$).  The
magnetic field and the field gradient directly above the edge can be
written as

\begin{equation}
B_{film}=\frac{\mu_0}{2\pi}\frac{hM}{z} \quad\mbox{and}\quad
B'_{film}=-\frac{\mu_0}{2\pi}\frac{hM}{z^2}.
\end{equation}

These expressions are analogous to those derived using Biot-Savart's
law for the magnetic field above an infinitely long and thin
current-carrying wire. The similarity between a permanent magnetic
film and a current-carrying wire can be explained using a simple
model. An unmagnetized film is comprised of many small magnetic
domains of random orientation.  The magnetic field produced by each
domain is equivalent to that from an imaginary surface current
flowing along the domain borders, perpendicular to the magnetization
vector \cite{Jac99}. For a uniformly magnetized film with
perpendicular anisotropy all domains are aligned in the same
direction (out of plane) and within the bulk the magnetic fields of
neighbouring domains cancel. A net effective current exists about
the perimeter of the film with a magnitude given by the product of
the magnetization and the film thickness ($I_{eff}~=~hM$).

The application of a uniform bias field ($B_{bias}$) in the $-x$
direction produces a radially symmetric two-dimensional quadrupole
magnetic field above the film edge at the height $z_0$, where the
magnitudes of $B_{bias}$ and $B_{film}$ are equal. To realize a
three-dimensional (3D) magnetic trap for weak-field seeking atoms a
nonuniform axial field $B_y$ is provided by two parallel currents
located beneath and perpendicular to the waveguide. Additionally,
$B_y$ suppresses spin-flip losses by preventing the total magnetic
field at the trap bottom from going to zero.  This results in a 3D
harmonic film trap at a distance $z_0$ from the surface with radial
frequency given by

\begin{equation}
2 \pi f_{radial} =
\frac{\mu_0}{2\pi}\frac{hM}{{z_0}^2}\sqrt{\frac{\mu_Bg_Fm_F}{mB_y}},
\end{equation}

where  $\mu_B$ is the Bohr magneton, $g_F$ is the Land\'{e} factor,
$m_F$ is the magnetic quantum number and $m$ is the atomic mass. The
ability to produce high quality, thick magnetic films with large
magnetization is necessary to produce tightly confining magnetic
traps for ultracold atoms.

\section{Tb$_6$Gd$_{10}$Fe$_{80}$Co$_4$ magneto-optical films and their properties}

The desire for large capacity information storage devices has
encouraged an extensive investment toward developing novel magnetic
compositions. These are primarily optimized to achieve small scale,
recordable patterning of magnetic media.  While it is possible to
benefit from this experience, applications with cold atoms have
several additional yet very specific requirements. Firstly, a high
Curie temperature ($T_{C}$) will prevent demagnetization during the
bake-out procedure, a necessary step in achieving UHV conditions.
Secondly, a high coercivity ($H_C$) will prevent the loss of
magnetization when applying large external magnetic fields. Finally,
the remanent magnetization ($M_{R}$) and the saturation
magnetization ($M_{S}$) should be large and nearly equivalent, an
indication of good magnetic homogeneity.  These conditions are
satisfied by Tb$_6$Gd$_{10}$Fe$_{80}$Co$_4$ magneto-optical films
which have a high Curie temperature ($T_{C}~\approx300~^{\circ}$C),
perpendicular anisotropy and a square hysteresis loop.

TbGdFeCo films were produced using a thin film deposition system
(Kurt J Lesker CMS-18) equipped with magnetron sputtering and
electron beam evaporation sources~\cite{Wan05}.  A composite target
with a nominal atomic composition of Tb$_6$Gd$_{10}$Fe$_{80}$Co$_4$
and a high purity chromium target are used in the production of the
magnetic films.  A systematic study of the influence of process
parameters over the properties of the film indicated that
deterioration of the magnetic anisotropy occurs for film thickness
above 250~nm. In order to maintain good magnetic properties and
increase the magnetic field strength near the surface we have
implemented a multilayer deposition which produces high quality
TbGdFeCo magnetic films with a total thickness approaching 1~$\mu$m.
A glass slide substrate was cleaned in an ultrasonic bath using a
nitric acid solution then carefully rinsed before being mounted in
the deposition chamber. The base pressure was less than
5$\times$10$^{-8}$~Torr prior to introducing the argon buffer gas
($\sim$4~mTorr). The substrate was then heated to 100~$^{\circ}$C
and a bonding layer of chromium (120~nm) was sputtered on the
surface. This was followed by the deposition of six bi-layers of
TbGdFeCo (150~nm) and Cr (120~nm) films.

\begin{center}
\begin{figure}
\begin{center}
\epsfbox{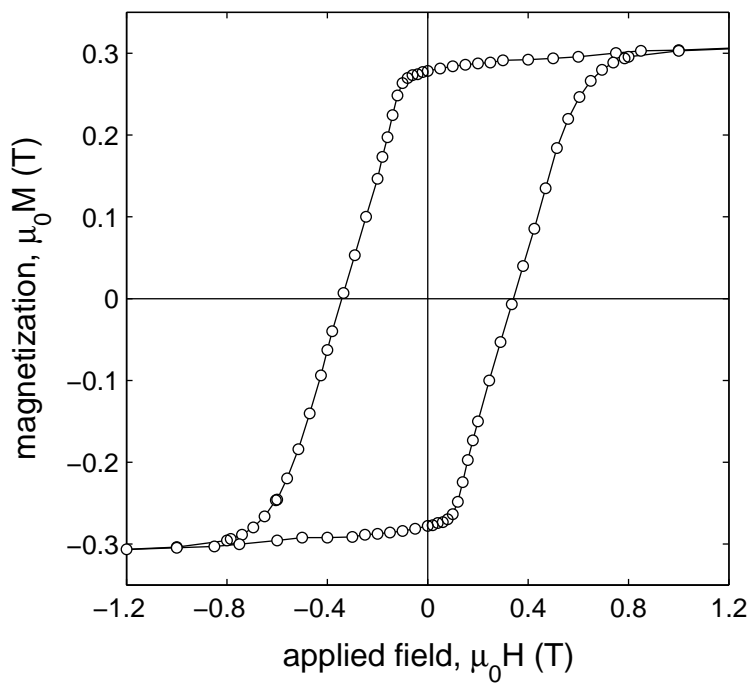}
\end{center}
\caption{\label{squid}A hysteresis loop derived from SQUID
magnetometry of a multilayer Tb$_6$Gd$_{10}$Fe$_{80}$Co$_4$ magnetic
film.  The film magnetization is $\mu_0M_S~\sim~\mu_0M_R~=$~0.28~T
and the coercivity is $\mu_0H_C~=$~0.32~T.}
\end{figure}
\end{center}

\begin{center}
\begin{figure}
\begin{center}
\epsfbox{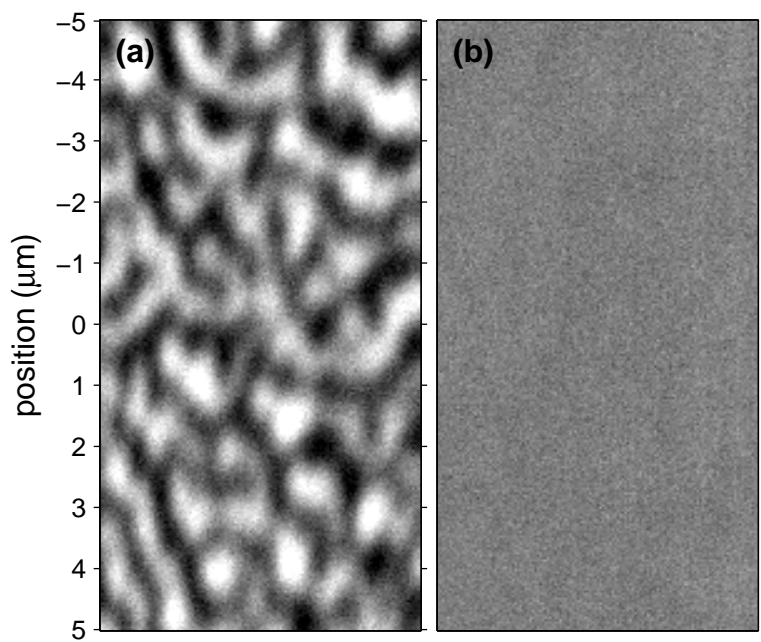}
\end{center}
\caption{\label{mfm}Magnetic Force Microscope (MFM) image of a
Tb$_6$Gd$_{10}$Fe$_{80}$Co$_4$ magnetic film surface.  (a)
Unmagnetized sample shows domain structure with micron sized
features. (b) Uniformly magnetized sample is free of any visible
magnetic structure.}
\end{figure}
\end{center}

The magnetic properties of the multilayer TbGdFeCo/Cr film were
characterized by a SQUID magnetometer (Figure~\ref{squid}).  The
hysteresis loop indicates a remanent magnetization of ~0.28~T for a
total magnet thickness of 900~nm ($hM~=~0.20\pm0.01$~A). Complete
magnetization of the film can be achieved by applying a field of
$\sim$~0.8~T, while the film magnetization is robust in the presence
of external fields below $\sim0.1$~T. The surface features of the
films have also been examined by a high-resolution atomic force
microscope operating in magnetic force mode (Figure~\ref{mfm}). An
unmagnetized sample shows micron-sized features consistent with
domain stripes, while an example of a uniformly magnetized sample
exhibits excellent magnetic homogeneity.

\section{Atom chip design}
\begin{center}
\begin{figure}
\begin{center}
\epsfbox{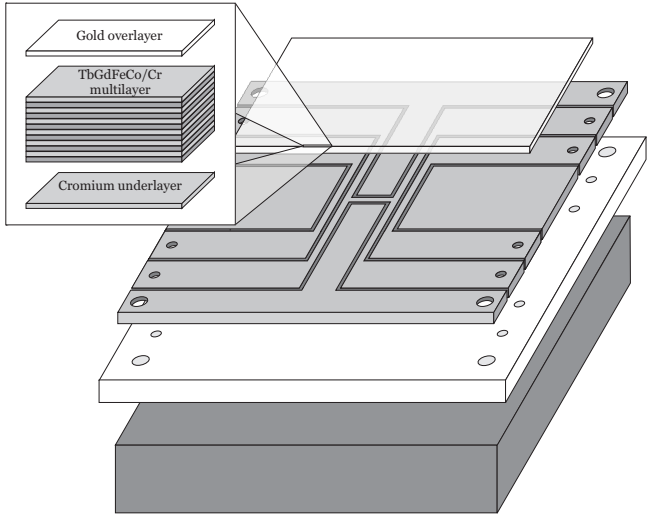}
\end{center}
\caption{\label{schem}Schematic view of the hybrid atom chip. Inset:
TbGdFeCo/Cr multilayer film and Au overlayer.  From the top down,
glass slide coated with magnetic film, machined silver foil H-wire
and end wires, Shapal-M base-plate and Cu heat sink.  Missing from
the schematic is the second glass slide and two rubidium dispensers}
\end{figure}
\end{center}
This device represents the first atom chip based on a
perpendicularly magnetized permanent magnetic film for trapping
ultracold atoms. It has been designed for the production and
manipulation of a BEC near the surface of the magnetic material.
Although these films are well suited for making tight and stable
trapping potentials up to a few 100~$\mu$m from the surface, the
small volume of the film trap is not suitable for efficient loading
directly from a magneto-optical trap (MOT).  To circumvent this
difficulty a current-carrying wire structure located beneath the
magnetic film provides an additional trapping field. The combination
of both the magnetic film and the wire structure represents the
hybrid atom chip design shown schematically in Figure~\ref{schem}.

The top layer of the atom chip consists of two adjacent 300~$\mu$m
thick glass slides which are sturdy enough to prevent warping.  The
long edges of the glass slides were polished with aluminium oxide
grit prior to deposition to remove visible chips. A multilayer
TbGdFeCo/Cr film was deposited on one slide using the procedure
outlined in Section 3. Both slides were then coated with a gold
overlayer (170~nm) and together form a large reflective surface
(40$\times$46~mm$^2$). This allows the collection of a large number
of atoms into a mirror MOT within a single-chamber UHV system.  The
glass slide coated with the TbGdFeCo/Cr multilayer film was then
magnetized in a uniform field of 1~T pending assembly.

The second layer of the hybrid chip is a wire structure which was
produced using the micro-machined silver foil technique developed by
Vale \textit{et al} \cite{Val04}. A 500~$\mu$m thick silver foil
(99.99 $\%$~purity) was fixed with epoxy (Epotek H77) to a 2~mm
thick Shapal-M machineable ceramic base-plate.  A computer
controlled Quick Circuit 5000 PCB mill was used to cut 500~$\mu$m
wide insulating grooves in the foil.  Each wire has a width of 1~mm
which is broadened to 6~mm far from the trapping region to
facilitate good electrical connections.  After cutting, the
insulating channels were filled with additional epoxy to increase
the structural integrity and thermal conductivity.  The wire
structure including electrical connections has a total resistance of
4.6~m$\Omega$.  A continuous current of 30~A can be applied with an
associated temperature rise of less than 40~$^{\circ}$C and
negligible increase in vacuum pressure.

In conventional atom chips U or Z-shape wires are used for creating
quadrupole and Ioffe-Pritchard ($IP$) magnetic field geometries to
realize mirror MOTs and magnetic microtraps \cite{Rei99}. In the
present atom chip, high currents are used to form a tight trap
relatively far from the wire, thereby avoiding unwanted collisions
with the surface of the slide. Consequently, the use of broad
conductors prohibit the use of separate U and Z-shape wires. This is
circumvented with a planar H-shape structure, designed to allow both
U and Z-shape current paths with good spatial overlap of the
associated traps. Axial confinement for the film trap is provided by
additional parallel conductors separated by 9.5~mm and located
either side of the H-shape structure. The top surface of the
machined silver foil was later polished flat to support the glass
slides.

During assembly, the polished edge of the TbGdFeCo film coated slide
is aligned to the middle of the H-shape structure and set with
epoxy. The second gold-coated slide is epoxied adjacent to the
magnetic film slide to complete the reflective chip surface. Two
rubidium dispensers are mounted on two ceramic blocks (Macor) which
are recessed below the chip surface. The two glass slides, machined
silver foil and ceramic base-plate are then fixed to a copper heat
sink. The completed chip is clamped to a 19~mm diameter solid copper
feedthrough (Ceramaseal, 800~A rating) and mounted in the vacuum
chamber. Electrical connections are made using 1.6~mm diameter bare
copper wire and BeCu barrel connectors in conjunction with a 12~pin
power feedthrough (Ceramaseal, 55~A rating).  A cold cathode gauge
indicated a pressure below 1$\times$10$^{-11}$~Torr after baking at
140~$^{\circ}$C for 4~days, highlighting the UHV compatibility of
all materials.

\section{Bose-Einstein condensation on a permanent magnetic film}

The reflective surface of the atom chip is used to form a mirror MOT
and accommodates 30~mm diameter laser beams provided by a high-power
diode laser (Toptica DLX110) locked to the D$_2$~($F~=~2\rightarrow
3$) cooling transition of $^{87}$Rb. The trapping light is detuned
18~MHz below resonance and has an intensity of 4~mW/cm$^{2}$ in each
beam. A repumping laser locked to the D$_2$~($F~=~1\rightarrow 2$)
transition is combined with the trapping light with an intensity of
0.5~mW/cm$^{2}$ per beam.  Two water-cooled coils mounted outside
the vacuum chamber provide a quadrupole magnetic field with gradient
0.1~T/m centered 4.6~mm below the chip surface. To load the mirror
MOT a current of 6.5~A is pulsed for 9.5~s through one resistively
heated Rb dispenser, allowing the collection of
$~$2$\times$10$^8$~atoms. The atoms are held for a further 15~s
while the UHV pressure recovers, ready for transfer to the
chip-based potentials.

Transfer begins by simultaneously ramping a current through the
U-shape circuit ($I_U$ = 0$\rightarrow$8~A), increasing the uniform
field $B_{bias}$ and turning off the external quadrupole magnetic
field over 50~ms.  This moves the atoms without loss, into a U-wire
MOT located 1.6~mm from the surface and increases the radial
gradient to 0.4~T/m. While this compression increases the spatial
overlap with the $IP$ potential, it also heats the cloud.  To
counteract this, the radial gradient is reduced rapidly to 0.11~T/m
with the trap light off to minimize any force on the atoms.
Polarization gradient cooling is applied for 2~ms with 56~MHz
red-detuned trap light to reduce the temperature from 140~$\mu$K to
40~$\mu$K. Both the MOT light and $I_U$ are then turned off leaving
the cold atoms in a uniform magnetic field.

Next a 200~$\mu$s optical pumping pulse is applied to maximize the
number of atoms in the $|F = 2, m_F = +2\rangle$ weak-field seeking
state ready for magnetic trapping.  A current ($I_Z$) of 21.5~A is
switched on through the Z-shape circuit while $B_{bias}$ is
increased to 1.3~mT to form an $IP$ wire trap at the same position.
A total of 4$\times$10$^7$~atoms are held with a background-limited
lifetime greater than 60~s.  Adiabatic compression of this trap is
performed by ramping $I_Z$ up to 31~A and $B_{bias}$ up to 4.0~mT
over 100~ms.  Further compression results in loss of atoms to the
surface.  The compressed magnetic trap is 560~$\mu$m from the film
surface where the radial and axial trap frequencies are
$2\pi\times$530~Hz and $2\pi\times$18~Hz respectively.  The elastic
collision rate in this trap ($\gamma_{el}~\approx~50$~s$^{-1}$) is
high enough to begin evaporative cooling.
\begin{center}
\begin{figure}
\begin{center}
\epsfbox{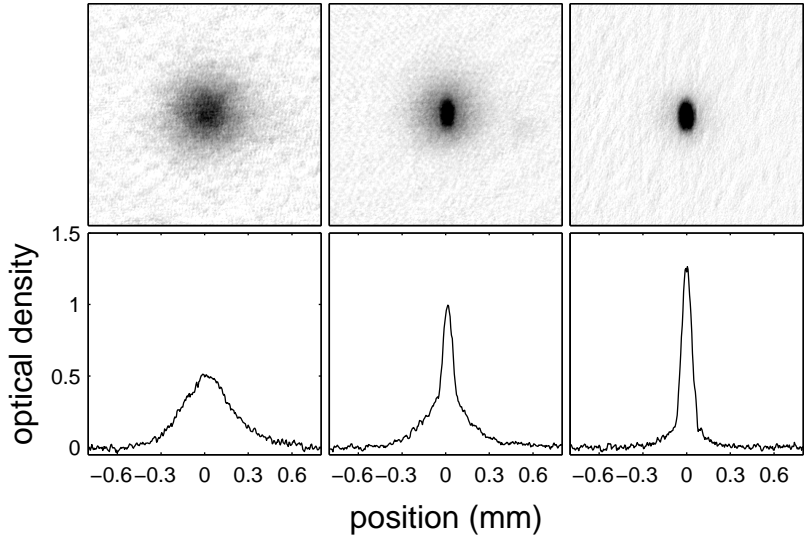}
\end{center}
\caption{\label{bec}Typical absorption images and optical density
profiles of a ballistically expanded atom cloud. Each image is a
single realization of the experiment where evaporation is performed
in the permanent magnetic film potential.  After truncating the
evaporation ramp, atoms are held for 150~ms and ballistically
expanded for 30~ms before imaging. (a) RF$_{final}$=804~kHz -
thermal cloud, (b) RF$_{final}$=788~kHz - partially condensed cloud,
(c) RF$_{final}$=760~kHz - an almost pure condensate.}
\end{figure}
\end{center}
Forced evaporative cooling to the BEC transition begins in the wire
trap and is then transferred to the film trap during a single
logarithmic radio frequency (RF) ramp. The first 8.85~s of this ramp
is a preliminary cooling stage in the wire trap down to a
temperature of $\sim~5~\mu$K. As the cloud is cooled the trap is
compressed further to improve the evaporation efficiency by lowering
I$_Z$ to 25 A, moving the trap to $350~\mu$m from the surface and
increasing the radial trap frequency to $\sim~2\pi\times$660~Hz. The
RF amplitude is then reduced to zero for 150~ms while the atoms are
transferred closer to the chip surface and finally to the film trap.
In this trap $I_Z$ is zero and axial confinement on the magnetic
film edge is provided by the two end wires, each with a current of
6~A. The trap bottom is tuned using an additional magnetic field
parallel to the film edge to minimize any discontinuity in the RF
evaporation trajectory. The radial and axial trap frequencies are
$2\pi\times$700~Hz and $2\pi\times$8~Hz respectively.  The RF
amplitude is then increased again and evaporation continues for 1~s
to the BEC phase transition.

Before imaging, the magnetic film trap is adiabatically moved
0.17~mm from the surface to avoid excessive field gradients from the
film. The cloud is then released by switching off $B_{bias}$ and the
atoms fall under gravity with minor acceleration from the permanent
field gradient. Resonant optical absorption is used to image the
atoms with a 100~$\mu$s $\sigma^+$ light pulse parallel to the gold
surface and tuned to the D$_2$~($F~=~2\rightarrow 3$) transition. A
CCD camera records the absorption image of the cloud using an
achromatic doublet telescope with a resolution of 5~$\mu$m/pixel.
Using the above procedure a new condensate of $1\times10^5$~atoms is
created every 50~s. Figure~\ref{bec} shows absorption images and
optical density profiles after 30~ms of ballistic expansion. The
forced RF evaporation is truncated at 804~kHz, 788~kHz and 760~kHz
revealing a thermal cloud, partially condensed cloud and nearly pure
condensate, respectively.

It is also possible to form a condensate trapped solely by the wire
trap. Here a single, uninterrupted, 10~s RF ramp results in a BEC
with atom number comparable to that realized in the film trap. This
provides a unique possibility for studying the properties of a BEC
in both permanent magnetic and current-carrying trapping
environments.  In addition, the formation of a BEC independent of
the top layer will allow new magnetic structures or materials to be
replaced with ease. The wire trap can also be used to transport a
BEC to regions on the chip where the magnetic field topology may be
different from those near the substrate edge.

\section{Magnetic field characterization}

The magnetic properties of the TbGdFeCo film were measured prior to
mounting on the atom chip using a combination of SQUID and magnetic
force microscopy.  In-situ techniques using cold atoms have been
employed to characterize the magnetic field produced by the film
inside the vacuum chamber.  This allows a direct comparison with the
simple model described earlier. A magnetically trapped cloud of cold
atoms or a BEC behaves as an ultra-sensitive probe to the local
magnetic field.  A measure of the trap position as a function of
$B_{bias}$ determines $B_{film}(z)$, while an independent measure of
 the trap frequency is used to determine $B'_{film}(z)$.

Once the BEC is confined by the film trap it is possible to profile
the magnetic field dependence near the surface. The potential
minimum is located at the point where the uniform magnetic field is
equal in magnitude to and cancels the field from the film ($B_{bias}
= -B_{film}$). The uniform magnetic field can be increased
(decreased) to move the trap minimum closer to (further from) the
film surface.  The BEC follows the potential minimum and the
measurement of cloud position with respect to the film surface
determines $B_{film}(z)$.  The strength of $B_{bias}$ is calibrated
within the vacuum chamber using untrapped atoms (far from the film)
and a short RF pulse resonant with the Zeeman splitting. The pixel
size in the imaging plane is calibrated against the gravitational
acceleration of freely falling atoms and agrees with the calibration
given by imaging a reference rule external to the apparatus.
 Unfortunately though, the glass substrate coated with magnetic material
 has recessed approximately 50~$\mu$m behind the second blank glass slide
as a consequence of unevenly cured epoxy.  The exact position of the
film surface (in relation to the image) is therefore unknown and
presents an uncertainty in $B_{film}(z)$.  For this reason a second
technique has been applied to provide more information about the
magnetic field from the film.
\begin{center}
\begin{figure}
\begin{center}
\epsfbox{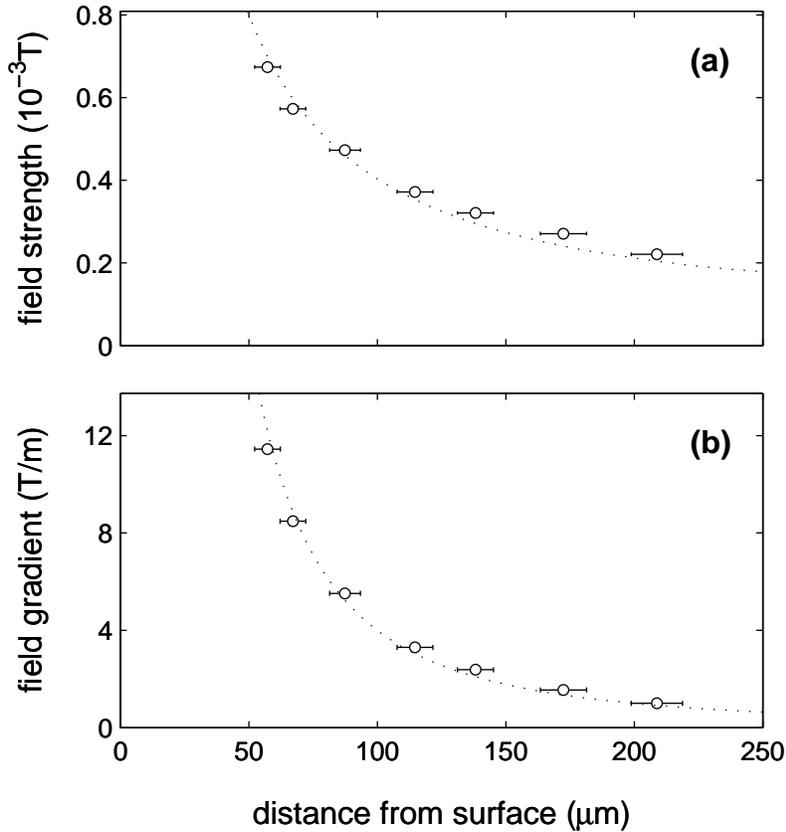}
\end{center}
\caption{\label{ieff}Measurements of the magnetic field strength (a)
and field gradient (b) as a function of distance from the surface.
The data (open circles) agrees well with predictions (dotted line)
of the simple model (see Equation 1). Experimental errors are mostly
determined by image resolution and a small uncertainty in the pixel
size calibration. }
\end{figure}
\end{center}
Harmonic oscillations with small amplitude and frequencies up to
10~kHz can be measured accurately over many periods with a BEC due
to low damping rates and small spatial extent. In this case, trap
frequencies are measured by exciting radial center of mass motion
within the film trap and have been measured to better than 1~Hz
($\sim$~0.1\% accuracy). These excitations were observed by rapidly
increasing the uniform magnetic field by approximately 5\% before
returning to the original position within 2 to 5~ms. The cloud
position was measured after 10~ms of free expansion and data has
been taken over five periods of oscillation.  In addition, the trap
bottom was measured using RF outcoupling with an accuracy better
than 10~mG ($\sim$~1\%). The measurement of trap frequency in
combination with the trap bottom ($B_y$) unambiguously determines
the local magnetic field gradient (see Equation 2).  This combined
with the trap position measurements have been used to provide the
magnetic field and the magnetic field gradient as a function of
height above the surface (Figure ~\ref{ieff}). This data is
consistent with a prediction based on the simple model where the
film thickness-magnetization product is given by the prior SQUID
measurement~($hM=0.20$~A).

\section{Discussion and conclusion}

We have demonstrated a hybrid atom chip that exploits
perpendicularly magnetized film or current-carrying wires for the
production of a BEC.  We have developed a multilayer magnetic film
structure (TbGdFeCo/Cr) that provides large magnetization and
thickness, important for realizing tight and flexible magnetic
microtraps.  We have used the BEC as a sensitive probe to directly
measure the local magnetic field and gradient associated with the
magnetic film. These measurements justify the use of the simple
model for perpendicularly magnetized magnetic microstructures.

At present we are extending the technique of cold atom magnetometry
to the measurement of the spatial dependence of the magnetic field
along the film edge.  Spatially dependent magnetic field variations
have been observed above microfabricated wire-based atom chips and
have been attributed to spatial deviations along the wire edge
\cite{Est04,Wan04}. Similar phenomena observed in permanent magnetic
structures may be caused by substrate roughness, deposition
irregularity or ultimately domain reversal.  Future studies are
aimed at the interaction between a BEC and magnetic thin films. A
comparison of the decoherence rates of condensates confined in
either the film or wire-based microtraps may reveal intriguing
possibilities for coherent manipulation of cold atoms in
microstructured permanent magnetic potentials.

\ack{ We would like to thank J.~Wang and D.~Gough for carrying out
the magnetic film deposition.  This project is supported by the ARC
Centre of Excellence for Quantum-Atom Optics and a Swinburne
University Strategic Initiative fund. }


\Bibliography{25}

\bibitem{Han01} H\"{a}nsel W, Hommelhoff P, H\"{a}nsch T W and J Reichel J 2001 {\it Nature} {\bf 413} 498-501

\bibitem{For01} Ott H, Fort\'{a}gh J, Schlotterbeck G, Grossmann A and Zimmermann C 2001 {\it Phys. Rev. Lett.} {\bf 23} 230401

\bibitem{Lea02} Leanhardt A, Chikkatur A, Kielpinski D, Shin Y, Gustavson T, Ketterle W and Pritchard D 2002 {\it Phys. Rev. Lett.} {\bf 89} 040401

\bibitem{Mul99} M\"{u}ller D, Anderson D Z,  Grow R J, Schwindt P D D and Cornell E A 1999 {\it Phys. Rev. Lett.} {\bf 83} 5194

\bibitem{Cal99} Calarco T, Hinds E A, Jaksch D, Schmiedmayer J, Cirac J I and Zoller P 1999 {\it Phys. Rev. A} {\bf 61} 022304

\bibitem{Wang05} Wang Y-J, Anderson D Z, Bright V M, Cornell E A, Diot Q, Kishimoto T, Prentiss M, Saravanan R A, Segal S R and Wu S 2005 {\it Phys. Rev. Lett.} {\bf 94} 090405

\bibitem{Shi05} Shin Y, Sanner C, Jo G-B, Pasquini T A, Saba M, Ketterle W and Pritchard D E 2005 {\it arXiv:cond-mat/0506464}

\bibitem{Sch05a} Schumm T, Hofferberth S, Andersson L M, Wildermuth S,
Groth S, Bar-Joseph I, Schmiedmayer J and Kr\"{u}ger P 2005 {\it
arXiv:quant-ph/0507047}

\bibitem{For02} Fort\'{a}gh J,  Ott H,  Kraft S,  G\"{u}nther A and Zimmermann C 2002 {\it Phys. Rev. A} {\bf 66} 041604

\bibitem{Est04} Est\`{e}ve J, Aussibal C, Schumm T, Figl C, Mailly D, Bouchoule I, Westbrook C I and Aspect A 2004 {\it Phys. Rev. A} {\bf 70}
043629

\bibitem{Jon03} Jones M P A, Vale C J, Sahagun D, Hall B V and Hinds E A 2003 {\it Phys. Rev. Lett.} {\bf 91} 080401

\bibitem{Sch05b} Scheel S, Rekdal P K, Knight P L and Hinds E A 2005 {\it arXiv:quant-ph/0501149}

\bibitem{Bar05} Barba I, Gerritsma R, Xing Y T, Goedkoop J B and Spreeuw R J C 2005 {\it Eur. Phys. J. D} 00055-3

\bibitem{Sin05a} Sinclair C D J, Curtis E A, Llorente Garcia I, Retter J A, Hall B V, Eriksson S, Sauer B E and Hinds E A 2005 {\it Preprint} arXiv:cond-mat/0503619

\bibitem{Sin05b} Sinclair C D J, Curtis E A, Llorente Garcia I, Retter J A, Hall B V, Eriksson S, Sauer B E and Hinds E A 2005 {\it Preprint} arXiv:physics/0502073

\bibitem{Eri04} Eriksson S, Ramirez-Martinez F, Curtis E A, Sauer B E, Nutter P W, Hill E W and Hinds E A 2004 {\it Appl.Phys. B} {\bf 79} 811

\bibitem{Jaa05} Jaakkola A, Shevchenkon A, Lindfors K, Hautakorpi M, Il'yashenko E, Johansen T H and Kaivola M 2005 {\it Eur. Phys. J. D} 00176-7

\bibitem{Jac99} Jackson J D 1999 {\it Classical Electrodynamics} 3rd Edition (New York: Wiley) chap 5

\bibitem{Wan05} Wang J Y, Whitlock S, Scharnberg F, Gough D S, Sidorov A I, McLean R J and Hannaford P 2005 {\it submitted J. Phys. D}

\bibitem{Val04} Vale C J, Upcroft B, Davis M J, Heckenberg N R and Rubinsztein Dunlop H 2004 {\it J. Phys. B: At. Mol. Opt. Phys.} {\bf 37} 29592967

\bibitem{Rei99} Reichel J, H\"{a}nsel W and H\"{a}nsch T D 1999 {\it Phys. Rev. Lett.} {\bf 83} 3398

\bibitem{Wan04} Wang D, Lukin M and Demler E 2004 {\it Phys. Rev. Lett.} {\bf 92} 076802









\endbib
\end{document}